\documentclass[amssymb, nobibnotes, aps, prl,twocolumn]{revtex4-1}
\usepackage{graphicx}
\usepackage{amsmath,mathrsfs,dcolumn,dsfont}
\usepackage{amssymb}
\usepackage{epsfig,epstopdf}
\usepackage{bm,bbm,mathrsfs}
\usepackage{threeparttable}
\usepackage{comment}
\usepackage{ulem}
\usepackage{tikz}
\usepackage[percent]{overpic}
\usepackage[large]{subfigure}
\usepackage{color}
\usepackage{xr}
\usepackage[top=30truemm,bottom=30truemm,left=25truemm,right=25truemm]{geometry}
\begin{document}


\title{Incoherent control of optical signals; quantum heat engine approach}

\author{Md Qutubuddin and Konstantin E. Dorfman}
 \email{dorfmank@lps.ecnu.edu.cn}
\affiliation{State Key Laboratory of Precision Spectroscopy, East China Normal University, Shanghai 200062, China}%

\date{\today}

\begin{abstract}
Optical pump-probe signals can be viewed as work done by the matter while transferring the energy between two coherent baths (from pump to probe). In thermodynamics a heat engine, such as laser, is a device which performs similar work but operating between two thermal baths. 
We propose an ``incoherent'' control procedure for the optical signals using the physics of quantum heat engine. By combining a coherent laser excitation of electronic excited state of molecule with thermal relaxation we introduce an effective thermal bath treating stimulated emission of probe photons as work performed by the heat engine. We optimize power and efficiency for the pump-probe signal using control parameters of the pump laser utilizing four level molecular model in strong and weak coupling regime illustrating its equivalence with the thermodynamic cycle of the heat engine.

\end{abstract}

\pacs{Valid PACS appear here}
\maketitle

The advancement in the field of quantum heat engines (QHE) attracted a lot of attention in the last  decade due to its connection with the real physical systems, such as lasers, solar cells \cite{sd11}, and biological systems \cite{dv13}. The maximum quantum efficiency for a three-level maser QHE, first introduced by Scovil and Schulz-DuBois \cite{ssd59,ssdap,ssd67} is governed by the Carnot bound obtained by invoking the detailed balance condition. Since then many theoretical proposals for thermodynamical heat engine in quantum regime have been discussed \cite{al79,prl98,pra74,ul15,agl7,lk17,sprl,cgg17}. In addition effects of profound quantum nature, such as quantum coherence and correlation and there influence on the performance of QHEs have been further investigated \cite{sz03,sd11,dv13}. Recent experiments  demonstrated that the QHE physics \cite{rd16, dd18, pb19} can be studied  using pump-probe optical measurements \cite{ kb17} where the working fluid is a radiation produced by resonantly driven electronic transitions in the material. The spectroscopic setup can be therefore viewed as the QHE which transfers energy from one heat bath (pump pulse) to another (probe pulse), while the work performed by the system is measured in the form of detected probe photons. Furthermore we employ the reservoir engineering \cite{pprl96, kqs99, cprl01} which has been previously studied in the context of thermodynamics \cite{zpra03, vnp09, ppra10, anj11, mepl11} as a control method for optical measurements utilizing the analogy between spectroscopic setups and quantum heat engines. While this analogy is not complete, since in QHE the system is in contact with thermal environment, while spectroscopic measurements are performed with coherent laser sources, there is a possibility to connect the two approaches in a sensible way. In particular, we introduce an effective thermal bath that mimics all the dynamical properties of the thermal bath by replacing a coherent laser excitation followed by a dissipative phonon-assisted relaxation due to internal degrees of freedom with an incoherent pumping. In the effective two-level system this can be achieved by matching the populations of the electronic states with two types of baths (coherent plus relaxation vs thermal). This consequently defines the range of parameters of the laser source which can be used to mimic thermal operation of the effective QHE. Once the laser parameters are fixed one can calculate the power and efficiency of the QHE assuming either strong or weak coupling between the molecular system and the probe field. The former represents a conventional QHE regime \cite{kpra18}, while the later is a typical case for spectroscopic measurements \cite{muknl}, where weak field-matter interactions allow the perturbative treatment of the signals. 


Various methods have been developed to optimize spectroscopic signals by carefully engineering  phase relations in the system-baths interactions. For instance, pulse shaping techniques allow to modify light absorption pathways using constructive and destructive interference by manipulating  phase of the optical pulses \cite{gos3}. Quantum control theory \cite{ys09} is yet another method, which gradually evolved from the studies of linear, closed, and small-scale systems to nonlinear, open, and large-scale networks. 
This optimization process relies on an elaborate numerical feedback loop algorithms. The utilization of the quantum control \cite{sb11} governs the system evolution through the Hamiltonian (unitary) dynamics and has therefore limited applicability for an open quantum systems. This limitation can be overcome by the addition of active manipulation of non-unitary (i.e., incoherent) evolution described by Liouville operator responsible for the effects of system environment. The control methods applicable to the dissipative (incoherent) dynamics due to the environment must be therefore developed. These new methods must be based on a different set of assumptions and are distinct from the coherent control methods applicable to the unitary system evolution \cite{pr06} as well as the thermal reservoir engineering \cite{lsr20} with feedback control \cite{mepj57,mpra04,spra06}.

\begin{figure}[t]
	\includegraphics[width=0.45\textwidth]{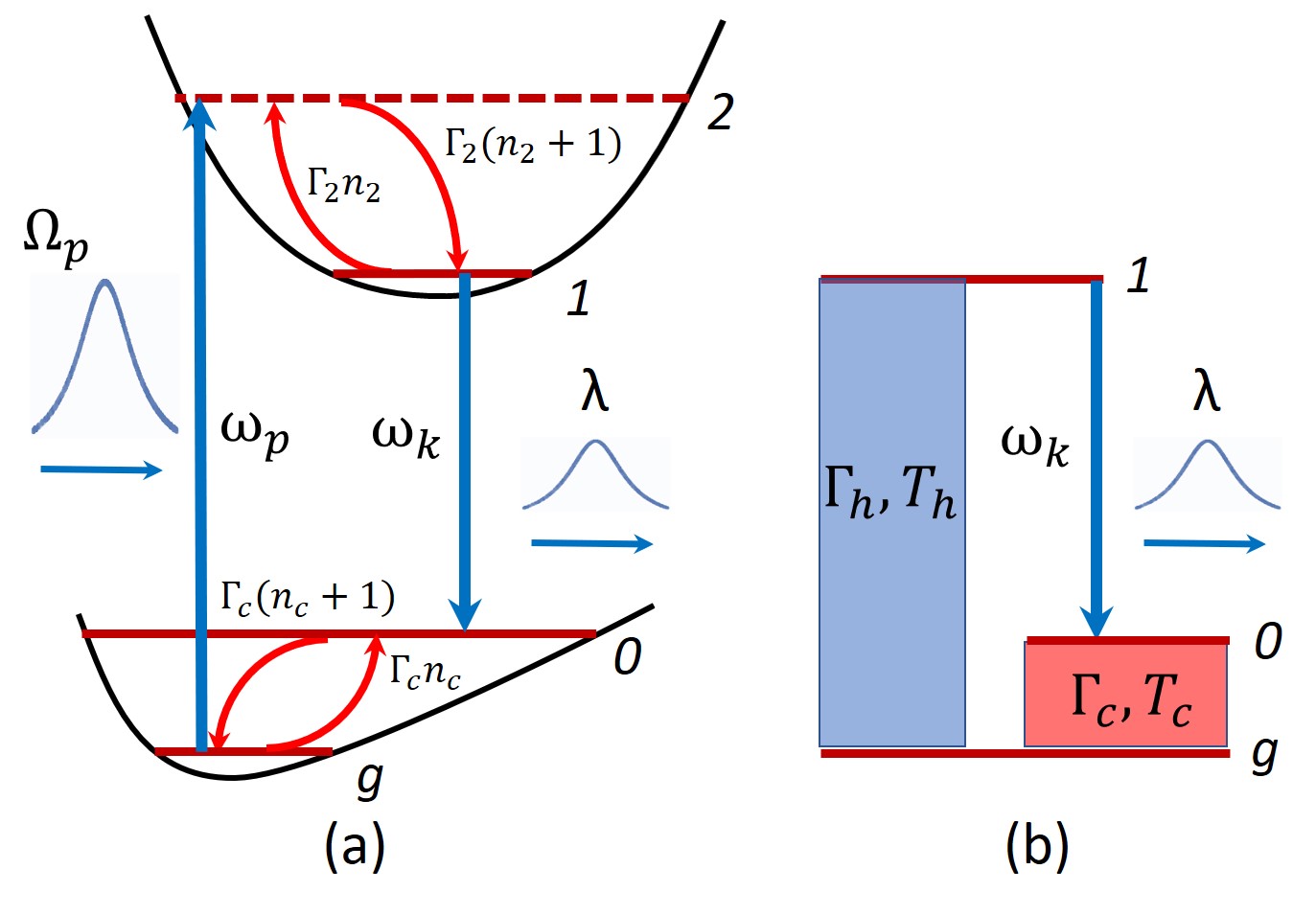}
	\caption{(color online)(a) Schematic of the pump-probe measurement in molecular system consisting of two electronic states.  Pump field resonant with electronic transition $g$-2 excites a vibrational wavepacket in the higher energy vibrational state 2 which relaxes to the lower energy vibrational state 1. Probe field then stimulates the emission from the state $1$ to the excited vibrational level $0$ of the  ground electronic state. Finally, vibrational relaxation brings the system back to its ground state $g$. (b)  Equivalent three-level QHE with energy levels transitions $g-1$ and $g-0$ driven by hot (at temperature $T_h$) and cold (at $T_c$) heat baths. The single-mode stimulated emission occurs at $1-0$ transition with the coupling strength $\lambda$.}\label{fig123}
\end{figure}

We consider a two-level molecular system with the ground state $g$ and the excited electronic state $e$ shown in Fig. \ref{fig123}a. We further consider two vibrational states of the excited electronic state $1$ and $2$. A coherent pump field excites resonantly the transition $g-2$ with Rabi frequency $\Omega_p$. The upper vibrational state 2 then relaxes to lower state 1 via the phonon emission.
The stimulated emission 1-0 via interaction with the probe field with Rabi frequency $\lambda$ followed by the thermal relaxation via interaction with the cold bath $0-g$ then brings the system to its initial ground state. The corresponding equation of motion for the density matrix is given by
\begin{eqnarray}\label{eq1L}
\dot{\rho}_{gg} &=& 2\Gamma_{c} [(n_{c} + 1)\rho_{00} - n_{c}\rho_{gg}]+ i \Omega_p (\rho_{g2} - \rho_{2g}), \notag\\
\dot{\rho}_{00} &=& -2\Gamma_{c} [(n_{c} + 1)\rho_{00} - n_{c}\rho_{gg}] + i \lambda (\rho_{01} - \rho_{10}), \notag\\
\dot{\rho}_{11} &=& \nonumber\Gamma_{2} [(n_{2}+1)\rho_{22} - n_{2} \rho_{11}] -  i \lambda (\rho_{01} - \rho_{10}) ,\notag\\
\dot{\rho}_{22} &=& - \Gamma_{2} [(n_{2}+1)\rho_{22} -  n_{2} \rho_{11}] - i\Omega_p(\rho_{g2} - \rho_{2g}), \notag\\
\dot{\rho}_{g2}&=& -[\Gamma_{2}(n_{2} + 1)/2 + \Gamma_{c} n_{c} - i(\omega_{2g} - \omega_{p})]\rho_{g2} \notag\\ &-& i\Omega_p(\rho_{gg} - \rho_{22}), \notag\\
\dot{\rho}_{2g}&=& - [\Gamma_{2}(n_{2} + 1)/2 + \Gamma_{c} n_{c} + i (\omega_{2g} - \omega_{p})]\rho_{2g} \notag\\ &+& i\Omega_p(\rho_{gg} - \rho_{22}),\notag\\
\dot{\rho}_{01}&=& -[\Gamma_{2} n_{2}/2 + \Gamma_{c} (n_{c}+1) - i(\omega_{10} - \omega_{k})]\rho_{01} \notag\\ &-& i\lambda(\rho_{11} - \rho_{00}), \notag\\
\dot{\rho}_{10}&=& - [\Gamma_{2}n_{2}/2 + \Gamma_{c} (n_{c}+1) + i (\omega_{10} - \omega_{k})]\rho_{10} \notag\\ &+& i\lambda(\rho_{11} - \rho_{00}),
\end{eqnarray}
where $\Gamma_2/2$ is the dephasing rate and $n_{2}=[\exp(\hbar \omega_{21} /k_BT_c)-1]^{-1}$ is the average phonon occupation number corresponding to $1 \leftrightarrow 2$ transition at ambient temperature $T_c$. Eqs. (1) has been derived using Born-Markov approximation assuming weak near-resonant pump field $\Omega_p\ll\Gamma_2n_2$. In this work we follow the pattern established in the series of earlier works which defined the framework for the laser QHE \cite{sz03, sd11, kpra18}, where we assume system in contact with the thermal phonon bath with relaxation described by a constant rate $\Gamma_2$. This represents a limit of the fast nuclear dynamics. Generally the connection between lineshape function $g(t)$ and the spectral density is given by \cite{muknl} $g(t)=\frac{1}{2\pi} \int_{-\infty}^{\infty} d\omega\frac{C(\omega)} {\omega^2} [\exp(-i\omega t) +i\omega t-1]$, where $C(\omega)$ is a spectral density function. Representing a bath by a collections of oscillators we obtain $C(\omega)=2\kappa \frac{\omega\Lambda} {\omega^2+\Lambda^2}$, where $\kappa$ is a reorganization energy, and $\Lambda$ is a nuclear dynamics timescale. Assuming nuclear dynamics to be fast compared to the coupling strength that governs magnitude of fluctuations $\hbar\Lambda^2\gg 2\kappa kT$ we obtain the homogeneous dephasing limit $Re[g(t)]=\Gamma_2 t$ such that $\Gamma_2=\kappa kT/(\hbar\Lambda)$ is the homogeneous dephasing limit. In more general case of spectral density, the solution of Eq. (1) can be found numerically as shown in \cite{kb17}.

We now investigate the four-level QHE based on Scovil and Schultz-Dubois maser \cite{ssd59} shown in Fig. 1b. Quantum heat engine regime is characterized by a strong coupling to the output radiation which makes it distinct from the spectroscopic regime, where probe field is typically weak. A hot reservoir at temperature $T_{h}$ of the system resonant with the $g-1$ transition and a cold reservoir at temperature $T_{c}$ coupled with the $g-0$ transition. The time evolution of the system in rotating frame is given by \cite{kpra18, eln02}
\begin{eqnarray}
\dot{\rho} = - \frac{i}{\hbar}[H_{0} - \bar{H} + V_{R}, \rho_{R}] + \mathcal{L}_{c}[\rho_{R}] + \mathcal{L}_{h}[\rho_{R}],
\label{eqh1}
\end{eqnarray}
where $H_{0} = \hbar\sum_{i=g,0,1}\omega_{i}|i\rangle \langle i|$, $\bar{H} = \hbar \omega_{g}|g\rangle\langle g| + \frac{\hbar\omega} {2}(|1\rangle \langle 1| - |0 \rangle\langle 0|)$ and interaction with the probe field is described by $V_{R} = \hbar\lambda(|1\rangle \langle 0| + |0\rangle \langle 1|) $, where $\lambda=\mu_{10} \mathcal{E}_{pr}$ is the probe-matter coupling which is strong compared to other relaxation processes, where $\mu_{10}$ is a transition dipole moment and $\mathcal{E}_{pr}$ is the classical amplitude of probe field. Master equation (\ref{eqh1}) contains two parts. The first, unitary part contains interaction with the coherent probe field and is governed by a commutator term. The last two terms governed by Liouville operator are describing interaction with the thermal reservoirs (see Appendix A). While the latter assumes weak field-matter interaction which is typical for the thermal radiation, the former has no assumptions about the strength of the probe field. Following the pioneering work on laser QHE \cite{ssd59} and more recent work of Scully and others \cite{sz03, sprl, sd11, dv13, kpra18} we adopt the strong coupling to the probe field which represents the so-called QHE limit which corresponds to e.g. cavity radiation of the laser. The Liouville operator for the system-bath interaction is given by   $\mathcal{L}_{w}[\rho]=\Gamma_{w}(n_{w}+1)[2|g\rangle\langle g|\rho_{l_{w}l_{w}} -|l_{w}\rangle\langle l_{w}|\rho - \rho |l_{w} \rangle \langle l_{w}| ] + \Gamma_{w}n_{w}[2|l_{w} \rangle \langle l_{w}|\rho_{gg} - |g\rangle \langle g|\rho - \rho |g\rangle\langle g| ],$ where $w = c$(cold) and $ h$(hot) baths and $l_{w}$ is $0$ and $1$ for the hot and cold baths, respectively. Note, that the strong field is generally defined with respect to the diagonal terms of the Hamiltonian (bare eigen energies), rather than Liouville operator (relaxation process). Furthermore, the QHE power, heat flux and other characteristics are defined via a trace operator, which is invariant with respect to the choice of the basis set.  In refs. \cite{sd11, sz03, sprl, dv13, kpra18, mprl10} it has been demonstrated how to find the system density matrix using bare system eigenstates. In other works \cite{djpc15, vnp09, ppra10, anj11} authors chose to use dressed states, while others \cite{xc16, agl7} consider strong coupling to the phonon bath and redefined master equation in polaron frame. 

Before proceeding to the QHE model which is based on the solution of the complete set of equations given by Eq. (\ref{eq1L}) we first introduce an effective heat bath. To that end we assume that the pump is relatively weak $\Omega_p\ll\Gamma_2n_2$ and the coupling to the probe field is much stronger than the coupling to the phonon bath that governs $2-1$ transition which itself is stronger than that of the bath driving $0-g$ transition: $\lambda\gg \Gamma_2n_2\gg \Gamma_cn_c$. The latter condition can be obtained in variety of molecular systems \cite{hb89}. Under these conditions one can eliminate the state $0$ from the total system of equations (\ref{eq1L}) and consider only three states such that the coherent excitation $g-2$ followed by a relaxation $2-1$.
The solution of this reduced system for the ground $g$ and lowest excited state $1$ populations read
\begin{align}
&\rho^{c}_{11}(t)=\mathcal{N}_c(t)(n_2+1)(1-e^{-\tilde{\Gamma}  t} )\quad \rho_{gg}^{c}=1-\rho_{11}^{c}, 
\label{eq3a}
\end{align}
where superscript $c$ indicates the coherent bath and $\tilde{\Gamma} = \Gamma _2 \left(n_2+1\right)/4-\sqrt{\Gamma _2^2 \left(n_2+1\right)^2/16-2 \Omega_p ^2}$. Normalization function $\mathcal{N}_c(t)=[1+2 n_2 +n_2e^{-\tilde{\Gamma}  t}]^{-1}$ ensures that the population of an effective two-level system consisting of states $g$ and $1$ is conserved (see Appendix B). 
In the high temperature limit $n_2\gg 1$, and,  assuming $\Omega_p\ll\Gamma_2n_2$ we obtain $\tilde{\Gamma} \simeq \frac{4\Omega_p^2}{\Gamma_2(n_2+1)}$. Generally, the condition $n_2\gg 1$ represents the high temperature limit. This is the optimum regime for operation of the QHE \cite{klr14}. For a room temperature “cold bath” $T_c=0.0259$ eV. High temperature limit is valid for various e.g. acoustic-like phonon modes, with the energy scale up to meV range.   

The combined effect of the coherent excitation $g\to 2$ followed by the phonon relaxation $2\to 1$ can be replaced by an effective thermal bath at temperature $T_h$ with the average photon number $n_h=[\exp(\hbar\omega_{1g}/k_BT_h)-1]^{-1}$ and dephasing $\Gamma_h$. In this case the state $2$ can be eliminated and the corresponding equation of motion for the populations of $g$ and $1$ read
\begin{align}
\dot{\rho}_{11}=-\Gamma_h[(n_h+1)\rho_{11}-n_h\rho_{gg}],\quad \dot{\rho}_{gg}=-\dot{\rho}_{11},
\end{align}
which yields the time-dependent solution:
\begin{eqnarray}
 \rho^{th}_{11}(t)=\mathcal{N}_{th}n_h(1 - e^{-\Gamma_h(1+2n_{h})t}),~\rho^{th}_{gg}=1-\rho^{th}_{11}, \ \
\label{eq4}
\end{eqnarray} 
where superscript $th$ indicates the thermal bath and the normalization $\mathcal{N}_{th}=[1+2n_h]^{-1}$. In order to match the solutions of the effective thermal bath in Eq. (\ref{eq4}) with that of a coherent bath in Eq. (\ref{eq3a}) the corresponding $n_h$ and $\Gamma_h$ must satisfy
\begin{align}n_{h}(t) = \frac{ n_{2} + 1 }{  n_{2}e^{-\tilde{\Gamma}t} - 1  },\quad \Gamma_h(t) = \frac{ 4\Omega_p^{2} }{ \Gamma_{2}(n_{2} + 1)( 2n_{h}(t) + 1)}. \label{eq:match}
\end{align} 
Eq. (\ref{eq:match}) describes the time-dependent parameters of the effective thermal bath, which yields a complete match between the two baths (coherent and thermal) at any given time (see Fig. 2b). More detailed description of the environment will result in more complicated relation which can be solved numerically \cite{xu18}. One can further provide an approximate relation between the bath parameters that is time independent and is easy to analyze. Assuming $n_{2} \gg 1$ and arbitrarily fixing the time: $t^{*} = \tilde{\Gamma} ^{-1} \log(n_{2}/2)$ we obtain
\begin{equation}
n^{*}_{h} \equiv n_{h}(t^{*}) = n_{2},\quad \Gamma_h^{*} \equiv \Gamma_h(t^{*}) = \frac{2 \Omega_p^{2}}{\Gamma_{2} n^{2}_{2}}.\label{eq:match1}
\end{equation}  
Parameters in Eq. (\ref{eq:match1}) results in the population dynamics shown in Fig. 2d. It yields a qualitatively good agreement at initial time ($\rho_{gg}(0)=1$, $\rho_{11}(0)=0$) as well as near the steady state  ($\rho_{gg}(\infty)=\rho_{11}(\infty)=1/2$) as seen in Fig. 2a. The error of $\sim 8\%$ associated with the choice of $t^*$ becomes apparent at intermediate times as shown in Fig. 2c, where a comparison with the exact numerical solution of Eq. (\ref{eq1L}) for the full four-level system  is presented. While the general expression in Eq. (\ref{eq4}) yields negative steady state value for $n_h$ when $t\to\infty$, the relevant population dynamics is determined by $\tilde{\Gamma}$  which is a small number. Therefore the choice of $t^{*}$ that is displayed in Fig. 2d corresponds to the value of time at which populations approach the steady state values, while $n_h$ is a growing function. Note, that the choice of $t^*$ near the steady state, can be used in the following thermodynamic analysis. Note, that the result of Eq. (\ref{eq:match}) - (\ref{eq:match1}) is applicable for a wide variety of parameters and is valid in the high temperature limit only when $n_2\gg 1$. In the low temperature limit, coherent bath creates a population inversion ($\rho_{11}^{c}(\infty)=1$, $\rho_{gg}^{c}(\infty)=0$), while the thermal bath yields weak excitation: ($\rho_{11}^{th}(\infty)=0$, $\rho_{gg}^{th}(\infty)=1$), which corresponds to the low efficiency regime and thus it won't be considered any further.

\begin{figure}[t]
	\includegraphics[width=0.48\textwidth]{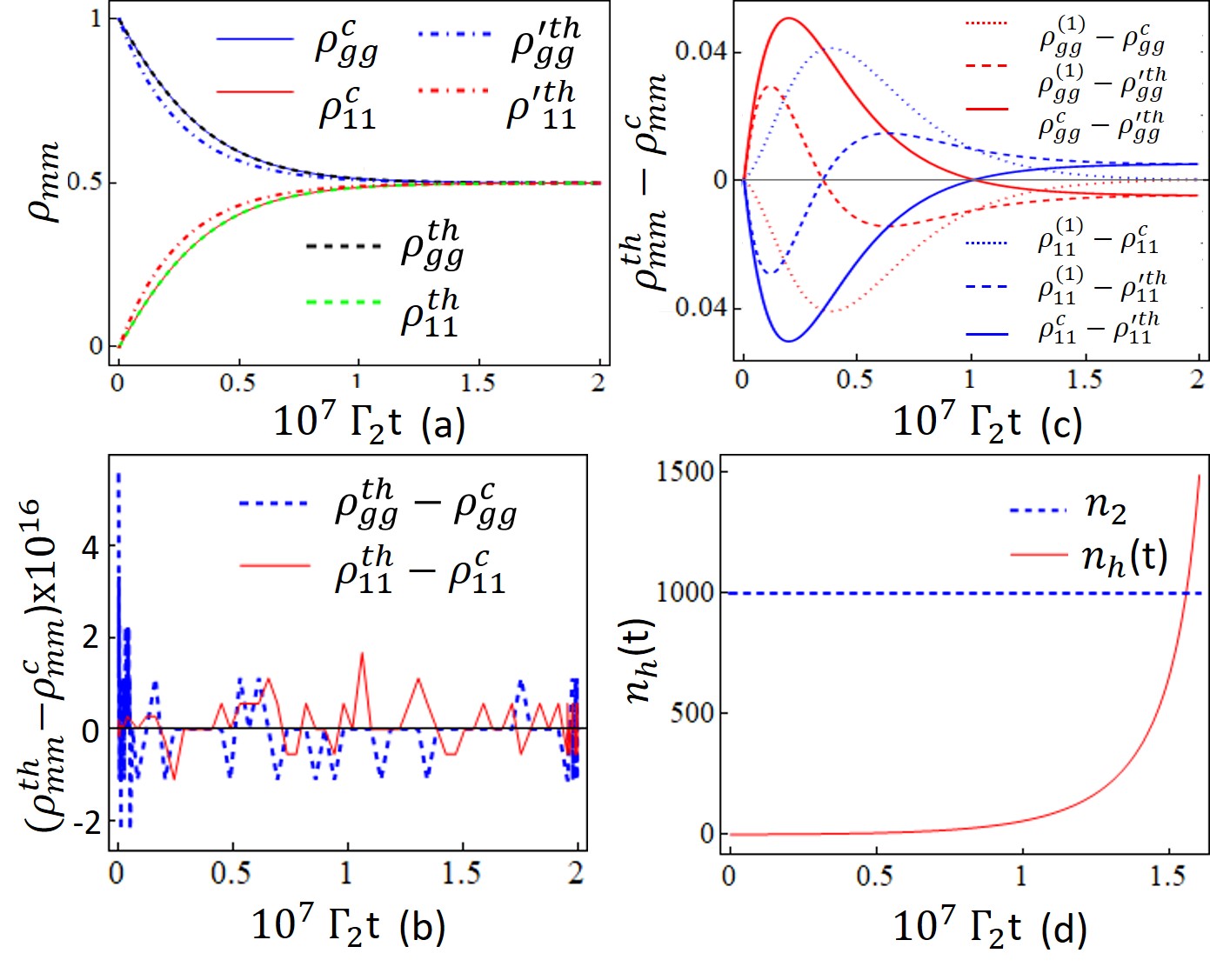}
	\caption{(Color online ) (a) The population of ground $g$ and lowest excited state $1$ obtained using coherent bath in Eq. (\ref{eq3a}) (solid lines) and thermal bath $\rho^{th}$ using parameters in Eq. (\ref{eq:match}) ($--$dashed) and $\rho^{'th}$ using Eq. (\ref{eq:match1})($\protect\tikz[baseline]{\protect\draw[line width=0.5mm,dash dot] (0,.4ex)--++(0.3,0) ;}$ dot-dashed). (b) The difference between  populations of coherent and thermal baths $\rho^c-\rho^{th}$ for parameters in Eq. (\ref{eq:match}). (c) same as (b) but for exact solution of Eq. (\ref{eq1L}) ($\rho^{(1)}$) vs approximate solutions of Eqs. (\ref{eq3a}) and (\ref{eq:match1}). (d) The time evolution of the microscopic occupation number $n_{h}(t)$ in Eq. (\ref{eq:match}) vs its approximation $n^{*}_{h} = n_{2}$ in Eq. (\ref{eq:match1}). The parameters for the simulations are $\Omega_p = 0.0001$ eV, $\Gamma_{2} = 0.025$ps$^{-1}$ and $n_{2} = 1000$.} \label{figc}
\end{figure}

The output power, and efficiency of a QHE described by Eq. (\ref{eq1L}) are given by \cite{prl98,pra74}  
\begin{align}
P^Q= -\frac{i}{\hbar}\text{Tr}\{[H_{0},V_{R}] \rho_{R}\},\quad \eta = -\frac{P}{\dot{Q}_h}, \label{eq:PQeta}
\end{align} 
where superscript $Q$ indicates the QHE power (not to be confused with spectroscopic power in Eq. (\ref{eq:Ps}), the heat flux is $\dot{Q_{h}}= \text{Tr}\{\mathcal{L}_{h}[\rho_{R}]H_{0}\}$.
Following the general approach outlined in \cite{kpra18} summarized in Appendix C we assume the high temperature limit and expand the occupation numbers  $n_h=n_2=T_c/\omega_{21}$, $n_c=T_c/\omega_c$ where $\omega_c=\omega_{0g}$. We then introduce an effective temperature of the hot bath $T_h=\Omega_p\omega_c(\Gamma_2\Gamma_c/2)^{-1/2}$, dimensionless temperature scale: $\tau=T_c/T_h$, pump energy scale: $c_p=\omega_p/\omega_c$, and coupling scale: $\lambda'=\lambda\omega_c/(\Gamma_cT_h)$. We then optimize the power in Eq. (\ref{genp}) with respect to dimensionless variable $\omega_{21}/\omega_c$ and obtain
\begin{align}
P_{max}^Q = P_0^Q\frac{2 \lambda^{'2} \{2 c_p \lambda^{'2} \tau^2 + c_{ p'}(\lambda^{'2} + \tau^2) - 2 \lambda' \tau \mathcal{C} \}}{3 (\lambda^{'2} - \tau^2)^2 },\label{gnpm}
\end{align}
where $P_0^Q=\Gamma_c\omega_c$, $c_{p'} =c_p - 1$, $\mathcal{C}=\sqrt{(\lambda^{'2} c_p + c_{p'} ) (c_p \tau^2 + c_{p'} )}$ and $\lambda' \ne \tau$. The efficiency corresponding to the maximum output power defined in Eq. (\ref{effa}) is thus given by
\begin{equation}
\eta^{*} =1-\left[c_p + \frac{\lambda^{'} \tau(\lambda^{'} \tau c_p-\mathcal{C})}{c_p \left(\lambda^{'2} + \tau^2 + 1\right) - 1} \right]^{-1}. \label{effg}
\end{equation}
Efficiency (\ref{effg}) is evaluated at fixed $\omega_c$ since this is a parameter of a given molecular system whereas $\omega_h$ can be manipulated via scanning of the pump frequency $\omega_p$.

\begin{figure*} 
\includegraphics[width=.95\textwidth]{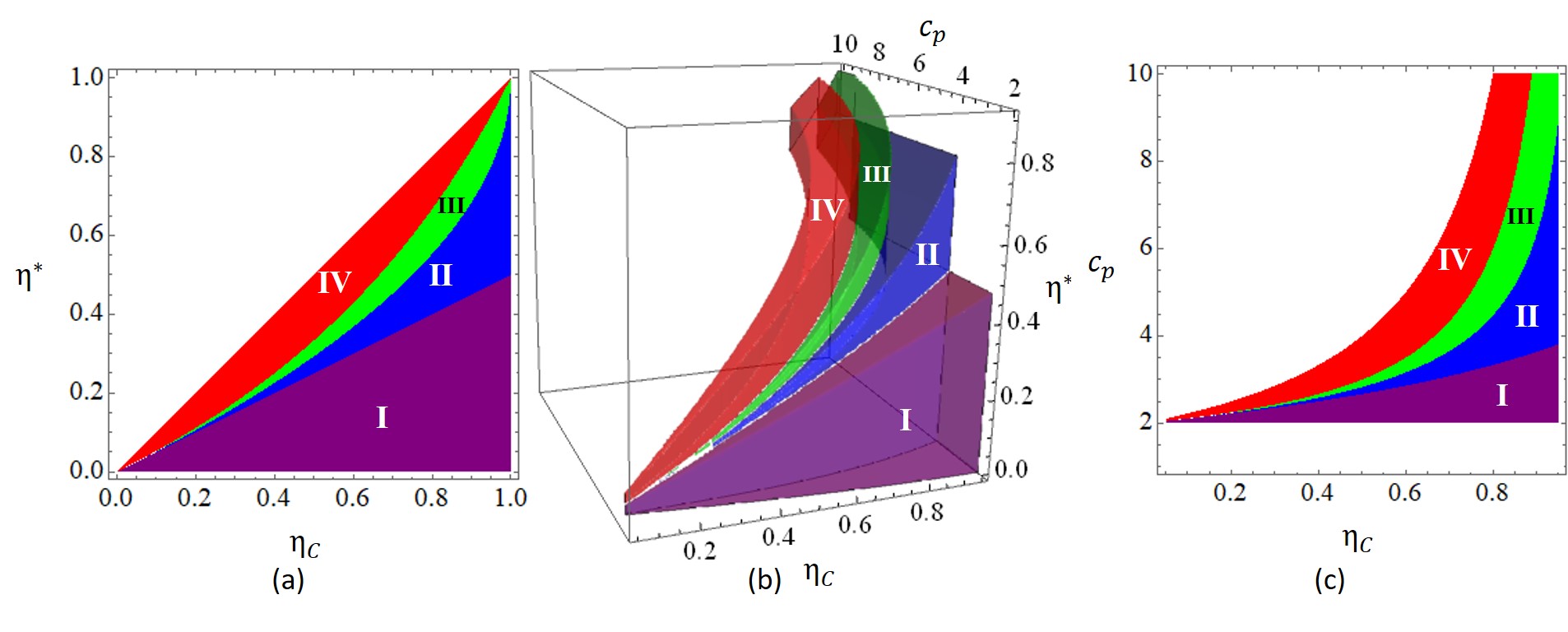}
\caption{(Color online) 2D mapping of the efficiency at maximum power $\eta^*$ in Eq. (\ref{efsc}) vs Carnot efficiency $\eta_C=1-\tau$ -(a) 3D mapping of $\eta^*$ vs $\eta_C$ vs dimensionless pump frequency $c_p$ - (b). 2D mapping of the $c_p$ vs $\eta_C$ corresponding to (a).} \label{fg5}
\end{figure*}

 \begin{table}
	\begin{tabular}{|c|c|c|}
		\hline\hline 
		Bound              & $\eta^*$ & $c_p$ \\
		\hline
		I & $0$ & $2$ \\
		I/II & $\eta_C/2$ & $4/(2-\eta_C)$   \\
		II/III& $\eta_{CA}$ &  $2/\sqrt{1-\eta_C}$ \\
		III/IV & $\eta_C/(2-\eta_C)$& $(2-\eta_C)/(1-\eta_C)$  \\
		IV&$\eta_C$ & $2/(1-\eta_C)$\\
			\hline
	\end{tabular}
	\caption{Parameters of the coherent bath corresponding to the QHE efficiency bounds shown in Fig. 3.}
\end{table}

We now compare the efficiency at maximum power (\ref{effg}) with the corresponding high temperature bounds of the QHE obtained in \cite{kpra18}. We first assume that the coupling with probe field $\lambda$ is the largest coupling in the system: $\lambda\gg \Gamma_hn_h,\Gamma_cn_c$ which yields
 \begin{equation}
 \eta^{*}_{SC} = 1-\left[c_p+\tau(\tau-\sqrt{1+\tau^2})\right]^{-1}, \label{efsc}
 \end{equation}
 where the subscript $SC$ indicates the strong coupling. Entire parameter space corresponding to the efficiency given by Eq. (\ref{efsc}) can be separated in four regions summarized in Table I represented by the colorful 2D shapes in Fig. 3a. The characteristic values describing the boundaries between the four regions correspond to $0$, $\eta_C/2$ (between I and II regions), $\eta_C/(2-\eta_C)$ (between III and IV), Carnot efficiency $\eta_C=1-\tau$ upper bound of IV) and Curzon-Ahlborn (CA) limit \cite{klr14} $\eta_{CA}=1-\tau^{1/2}$ (between II and III). Using dimensionless pump frequency $c_p$ as a control parameter which depends on the effective temperature ratio $\tau$ one can obtain the corresponding three dimensional parameter space for each of the regions that include $\{\tau,c_p,\eta^*\}$ as shown in Fig. 3b. Note, that the two parameters of the pump field: frequency $\omega_p$ and the Rabi frequency $\Omega_p$ which defines an effective hot bath temperature $T_h$  can be controlled experimentally.  Thus 2D parameter space $\{\tau,c_p\}$ shows a constrained relation between the two as seen in Fig. 3c. For instance, CA limit is obtained when $c_p\simeq 2/\sqrt{\tau}$. This corresponds to the Rabi frequency $\Omega_p^{(CA)}\simeq\omega_p^2\sqrt{\Gamma_c\Gamma_2/2}T_c/(4\omega_c)$. Similarly one can analyze other bounds corresponding to the standard QHE model \cite{kpra18}. There exists an additional constraint corresponding to parameter $\gamma=\Gamma_h/\Gamma_c$. For instance, CA limit is obtained when $\gamma\to 0$ which corresponds to $ \Omega_p\ll \sqrt{\Gamma_c\Gamma_2}n_2$. In another limit $\eta^*=\eta_C/(2-\eta_C)$  corresponds to $\gamma\to\infty$ which yields $ \Omega_p\gg \sqrt{\Gamma_c\Gamma_2}n_2$. 
 Therefore, by setting the specific relation between the pump frequency and intensity one can control energy conversion between the pump and the probe pulse near the maximum of the corresponding thermodynamic cycle. 
 Note, that one can exceed not only standard efficiency at maximum power limits such as CA, but also the Carnot limit. This is not a surprising result, since in the strong coupling limit QHE is approaching lasing threshold, which corresponds to the gain regime and the input coherent drive is not at thermodynamic equilibrium. Thus the system under coherent drive cannot be generally treated as a closed system. Therefore, Eq. (\ref{eqh1}) describes the system driven by the thermal pump and is therefore bounded by the Carnot limit applicable as long as $c_p<2(1-\eta_C)^{-1}$ according to regime IV in Table I.

So far we analyzed the case of the strong coupling to the probe field ($\lambda'\gg 1$), which corresponds to the laser QHE regime. We now focus on the regime typical for spectroscopic signals derived using perturbative expansions over light-matter interactions. 
We derive the pump-probe signal to second order in the pump as well as in the probe described by the double sided Feynman diagrams shown in Fig. \ref{fey} of Appendix D. Diagrams a and b illustrate how the two interactions with the pump pulse excite molecular system from its ground state $g$ to the excited state population $2$ which then relaxes to $1$ via phonon scattering. The consequent two interactions with the probe bring the system to $0$ state population. In diagram c the system remain in the ground state population $g$ after a Raman process initiated by the pump ($g-2-g$). Interaction with a cold bath promotes the system to the population of $0$ state and the consequent Raman process initiated by the probe ($0-1-0$) brings it back to the $0$ state population. The corresponding power is defined as a rate of change of the probe photon number multiplied by a photon energy $P_s=\omega_{10}\frac{d}{dt}\langle \hat{E}_{pr}^{\dagger}(t)\hat{E}_{pr}(t)\rangle$, which takes the form 
\begin{align}\label{eq:Ps}
P^s = \omega_{10}\mathcal{E}_{pu}^{2} \mathcal{E}_{pr}^{2} \text{Im}[R (-\omega_{pr}, \omega_{p})],
\end{align} 
where superscript $s$ indicates spectroscopic regime, $\mathcal{E}_{pu}$ and $\mathcal{E}_{pr}$ are the classical amplitudes of the pump and probe fields, respectively. Note, that both definitions of the output power in Eq. (\ref{eq:PQeta}) and (\ref{eq:Ps}) are equivalent and represent the rate of change of the probe photon energy flux according to Ref. \cite{sd11}. However, while Eq. (\ref{eq:PQeta}) contains definition originated from the general solution of the equations of motion for an arbitrary pump and probe intensities, Eq. (\ref{eq:Ps}) results from the perturbative expansion with respect to pump- and probe-matter interactions. The molecular response function has three terms $R(-\omega_{pr}, \omega_{p})= \sum_{j=a,b,c}R_{j} (-\omega_{pr}, \omega_{p})$ defined by Eqs. (\ref{R1}) - (\ref{R3}) of Appendix D corresponding to the three diagrams in Fig. \ref{fey}  \cite{muknl,mcr9}. Following a similar optimization of the power defined by Eq. (\ref{d12}) with respect to $\omega_{21}$ that results in Eq. (\ref{gnpm}) we obtain for the maximum power
\begin{align}\label{eq:Pmaxs}
P_{max}^s = \frac{\lambda^{'2} \gamma_cP_0^s}{f^2\tau^4}(f+2\sqrt{\gamma_c})^2(c_{p'}+\gamma_c+f\sqrt{\gamma_c}/2),
\end{align} 
where $P_0^s=6\Gamma_c^2T_c^2/(\sigma_p\omega_c)$, $\sigma_p$ is a bandwidth of the pump field, $f = \sqrt{\gamma_c}-\sqrt{8 c_{p'} + 9 \gamma_c} $, and $\gamma_c=\Gamma_2/\Gamma_c$. Since the spectroscopic signal represents power rather than the efficiency we will be comparing maximum power $P_{max}^Q$ in Eq. (\ref{gnpm}) vs $P_{max}^s$ in Eq. (\ref{eq:Pmaxs}). We first note, that while $P^s$ has been obtained to second order in the probe field, the QHE power $P^Q$ in Eq. (\ref{gnpm}) is derived to all orders in the probe. Therefore, we expand $P^Q$ to second order in $\lambda'$. Second, $P_{max}^s$ depends on extra parameter $\gamma_c$ which is absent in the QHE result. Maximized powers become equivalent $P_{max}^{Q}=P_{max}^{s}$ at $\gamma_c= 2c_p^2/(9\tau^2)$ assuming for brevity that $P_0^Q=P_0^s=1$. On the other hand the maximum of  Eq. (\ref{eq:Pmaxs}) with respect to $\gamma_c$ is achieved at $\gamma_c^*=c_{p'}/2$. Thus, Eqs. (\ref{gnpm}) and (\ref{eq:Pmaxs}) reduce to
\begin{align}\label{eq:PQS}
P_{max}^{Q*}=\frac{2\lambda^{'2}c_{p'}}{3\tau^2},\quad P_{max}^{s*}=\frac{\lambda^{'2}c_{p'}^2}{16\tau^4}.
\end{align}
To further compare expressions in (\ref{eq:PQS}) one can identify the pump pulse parameters ($\omega_p$) and $\Omega_p$ corresponding to limiting cases. For instance $P_{max}^{Q*}=P_{max}^{s*}$ at $c_{p'}^{*}=32\tau^2/3$ which corresponds to $\Omega_p^*=4T_c[\Gamma_2\Gamma_c/(3\omega_c\omega_p)]^{1/2}$. Furthermore, by setting  $c_{p'}^{**}=16\tau^2/3$ (i.e. $\Omega_p^{**}=\Omega_p^*/\sqrt{2}$)  we obtain $P_{max}^{Q**}=2P_{max}^{s**}=32/9$ which corresponds to the maximum of the difference between the two expressions. Thus, in the weak coupling regime one can also achieve the degree of control over the pump-probe signal by identifying the constrained parameter space for the pump field.

In summary, we have developed a novel method for control of the optical signals based on the analogy with quantum heat engines where  energy transfer occurs from the pump to the probe field. We found that the yield of the spectroscopic measurement can be improved when the corresponding regime is close to the thermodynamic cycle.  The proposed model can provide a reasonable qualitative guidance for the experimental implication in molecular systems consisting of two electronic states where the process of coherent excitation and consequent relaxation can be viewed as an effective thermal heat bath environment. This apparent connection between the thermodynamics of the QHE and the spectroscopy may emerge as a novel ``incoherent control'' tool for optimization of the optical measurements, which can enhance the yield of the fluorescence, the pump-probe measurements, and improve the signal-to-noise ratio in a wide class of the optical signals. The rich physics of the QHE can be extended to the Otto-cycle engines mimicking the time-domain nonlinear optical signals which will be the subject of the future studies. \\

We gratefully acknowledge the support from the National Science
Foundation of China (No. 11934011), the Zijiang Endowed Young
Scholar Fund, the East China Normal University and the Overseas
Expertise Introduction Project for Discipline Innovation (111 Project, B12024). MQ thanks P. Saurabh for valuable discussion and acknowledges the support from CSC Scholarship (CSC No. 2018 DFH 007778). \\

\setcounter{equation}{0}
\renewcommand{\theequation}{A\arabic{equation}}
\begin{center}
\textbf{\large{Appendix A: Derivation of the dissipative Lindblad superoperator }}
\end{center}

We consider a four-level system interacting resonantly with the classical pump and probe fields. Four-level system is in contact with the ambient reservoir at room temperature $T_c$. This thermal reservoir drives the relaxation between states 2-1 and 0-g. The Hamiltonian for the system-bath interaction for the transition $2-1$ reads
\begin{eqnarray}
\hat{V}(t) = \hbar \sum \text{g}_{k} \hat{b}_{k}e^{i(\omega - \nu_{k})t} |2 \rangle\langle 1| + \text{H.c.,} \label{s3}
\end{eqnarray}
and the Hamiltonian for the cold bath interacting with $0-g$ transition is given by 
\begin{eqnarray}
\hat{\mathcal{V}}(t) = \hbar \sum \text{G}_{q} \hat{b}_{q}e^{i(\omega - \nu_{q})t} |0 \rangle\langle g| + \text{H.c.,} \label{s4}
\end{eqnarray}  
where the system-bath coupling constant $\text{g}_{k} = \mathcal{P} _{21}.\hat{\epsilon}_{k} \mathcal{E}_{k}/\hbar$ and $\text{G}_{q} = \mathcal{P} _{og}.\hat{\epsilon}_{q} \mathcal{E}_{q}/\hbar$ with the dipole moment for transition $2 \leftrightarrow 1$ and $0 \leftrightarrow g$ is given by $\mathcal{P}_{21}$ and $\mathcal{P}_{og}$, respectively, and the polarization of the field is denoted as $\hat{\epsilon_{k}}$ and $\hat{\epsilon_{g}}$, respectively. The electric field per photon is $\mathcal{E}_{k} = (\hbar \nu_{k}/2\epsilon_{0}V_{ph})^{1/2}$ and $\mathcal{E}_{q} = (\hbar \nu_{q}/2\epsilon_{0}V_{ph})^{1/2}$ respectively, where $V_{ph}$ is the photon quantization volume. We assume that the system interacts with the reservoir represented by the reservoir density operator $\rho_{R}$. The equation of motion for the system density operator $\rho$ is given by
\begin{eqnarray}
\mathcal{L}_{21}[\rho] &=& -\frac{i}{\hbar}\text{Tr}_{R} \left [V(t),\rho(t_{0})\otimes \rho_{R} \right ]\nonumber\\
&-&\frac{1}{\hbar^{2}}\text{Tr}_{R}\int_{t_{0}}^{t} \left[ V(t) \left[ V(t'), \rho(t') \otimes \rho_{R}(t_{0}) \right]\right] dt'. \label{s5} \ \ \ \ \ \
\end{eqnarray} 
We note the bath operator represents the thermal state, i.e. $ \langle \hat{b}_{\textbf{k}} \rangle=\langle \hat{b}^{\dagger}_{\textbf{k}} \rangle = 0, \langle \hat{b}_{\textbf{k}}\hat{b}_{\textbf{k}'} \rangle = \langle \hat{b}^{\dagger}_{\textbf{k}} \hat{b}^{\dagger}_{\textbf{k}'} \rangle = 0, \langle \hat{b}^{\dagger}_{\textbf{k}} \hat{b}_{\textbf{k}'} \rangle = \bar{n}_{\textbf{k}} \delta_{\textbf{kk}'} $ and $\langle \hat{b}_{\textbf{k}} \hat{b}^{\dagger}_{\textbf{k}'} \rangle = (\bar{n}_{\textbf{k}}+1)\delta_{\textbf{kk}'}$ where $\bar{n}_\mathbf{k}=[\exp(\beta_c\epsilon_{\mathbf{k}})-1]^{-1}$ is the phonon occupation number corresponding to the cold temperature $T_c=\beta_c^{-1}$. Inserting $\hat{V(t)}$ in Eq. (\ref{s5}), we obtain
\begin{eqnarray}
\mathcal{L}_{21}[\rho]&=&-\frac{1}{\hbar^{2}}\int_{t_{0}}^{t}dt' \sum _{kk'}^{} \text{g}_{k}\text{g}_{k'} \big\{ (\text{X}_{+}\text{X}_{+}\rho(t') - 2 \text{X}_{+} \nonumber\\
&&\rho(t') \text{X}_{+} + \rho(t')\text{X}_{+}\text{X}_{+} ) \langle b_{k}b_{k'}\rangle e^{i(\omega - \nu_{k})t +} \nonumber\\ 
&& {}^{ i (\omega - \nu_{k'})t'} + (\text{X}_{+}\text{X}_{-}\rho(t') - \text{X}_{-}\rho(t')\text{X} _{+})\langle b_{k}b^{\dagger}_{k'}\rangle \nonumber\\ && e^{i(\omega - \nu_{k})t - i(\omega - \nu_{k'})t'} + (\text{X}_{-}\text{X}_{+}\rho(t') - \text{X}_{+}\rho(t')  \nonumber\\ && \text{X}_{-}) \langle b^{\dagger}_{k}b_{k} \rangle e^{-i(\omega - \nu_{k})t + i(\omega - \nu_{k})t}\big\} + \text{H.c.},\label{s61}
\end{eqnarray}
where $X_{+}= |2\rangle\langle 1|$ and $X_{-} = |1 \rangle\langle 2|$. On substituting the various expectation values from above paragraph in Eq. (\ref{s61}), we obtain 
\begin{eqnarray}
\mathcal{L}_{21}[\rho] &=& -\frac{1}{\hbar^{2}}\int_{t_{0}}^{t} dt' \sum_{k}^{}\text{g}^{2}_{k} \big\{ n_{k}\big(|1 \rangle\langle 1| \rho(t') + \rho(t') \nonumber\\
&& |1 \rangle\langle 1| - 2|2 \rangle\langle 2| \rho_{11}\big)e^{-i(\omega - \nu_{k})(t-t')} + (n_{k}+1)\nonumber\\
&& \big(|2 \rangle\langle 2|\rho(t') + \rho(t')|2 \rangle\langle 2| - 2|1 \rangle\langle 1|\rho_{22} \big)\nonumber\\ && e^{i(\omega-\nu_{k})(t-t')}\big\}.\label{s51}
\end{eqnarray}
The sum over $\textbf{k}$ is replaced by an integral through prescription 
\begin{equation}
\sum_{\textbf{k}}^{} \rightarrow \frac{V_{ph}}{\pi^{2}} \int_{0}^{\infty}dk k^{2}. \label{s6}
\end{equation}
We can assume that the density matrix varying slowly with the time and using the Markov approximation $\rho(t') \approx \rho(t)$. We then extend the upper limit of integration to infinity. Then the time integration yields
\begin{equation}
\int_{t_{0}}^{\infty} dt' e^{i(\omega_{2} - \nu_{k})(t-t')} = \pi \delta(\omega - \nu_{k}) + i\text{P}\frac{1}{\omega - \nu_{k}}, \label{s7}
\end{equation}

where $\text{P}$ denotes the Cauchy principal value. The above expression split the correlation into real and imaginary part to obtain the decays rate and Lamb shift, respectively. So, the Cauchy part will not affect our result and for simplicity, we consider the transition between the atomic levels only.
In the Weisskopf-Wigner approximation (Born-Markov plus RWA approximations), we replace $\text{g}_{k} \approx \text{g}_{k0}$ and $n_{k} \approx n_{k0}=n_{2}$ assuming broadband coupling, and obtain the final form of the Liouville operator driving 2-1 transition
\begin{eqnarray}
\mathcal{L}_{21}[\rho] &=& \gamma_{2}(n_{2}+1)\big(|1 \rangle\langle 1|\rho_{22} - \frac{|2 \rangle\langle 2| \rho + \rho |2 \rangle\langle 2|}{2} \big)\nonumber \\
&+& \gamma_{2}n_{2} \big( |2 \rangle\langle 2| \rho_{11} - \frac{|1 \rangle\langle 1|\rho + \rho |1 \rangle\langle 1|} {2} \big), \label{s8}
\end{eqnarray}
where $\gamma_{2} = \frac{2k_{0}V_{ph}\text{g}^{2}_{k0}}{\pi c}$. Similarly we derive the $\mathcal{L}_{c}[\rho]$ describing interaction with cold bath of the 0-g transition, which yields 
\begin{eqnarray}
\mathcal{L}_{c}[\rho] &=& \gamma_{c}(n_{c}+1)\big(2|g \rangle\langle g|\rho_{00} - |0 \rangle\langle 0| \rho - \rho |0 \rangle\langle 0| \big)\nonumber \\
&+& \gamma_{c}n_{c} \big( 2 |0 \rangle\langle 0| \rho_{gg} - |g \rangle\langle g|\rho - \rho |g \rangle\langle g| \big), \label{s9}
\end{eqnarray}
where $\gamma_{c} = \frac{4q_{0}V_{ph}\text{g}^{2}_{q0}}{\pi c}$, has an extra factor of $2$ compared
with $\gamma_2$ to be consistent with the earlier works \cite{kpra18}. Using equations (\ref{s8}) and (\ref{s9}), we obtain Eq. (\ref{eq1L}).

\setcounter{equation}{0}
\renewcommand{\theequation}{B\arabic{equation}}
\begin{center}
\textbf{\large{Appendix B: Effective thermal bath}}
\end{center}

The two electronic energy levels of the molecule shown in Fig. \ref{fig123}. Two vibrational states of the electronic ground state $g$ and $0$ while $1$ and $2$ are two vibrational state of the electronically excited state. Coherent pump pulse excites the transition $g-2$ with Rabi frequency $\Omega_p$. The corresponding equation of motion for the density matrix is given in Eq. (\ref{eq1L})(coherent bath) and steady state solution for the populations of  $g$, $1$ and $2$ states are given by
\begin{eqnarray}
	\rho_{gg}^{'c(ss)} = \rho_{22}^{'c(ss)} = \frac{n_2}{3n_2+1},\ \ \rho_{11}^{'c(ss)} =\frac{n_2+1}{3n_2+1}. \ \ \  \label{rcss}
\end{eqnarray}
In order to eliminate the state $2$ we normalize the above solution for the population of states $g$ and $1$ such that $\rho_{gg}^{c(ss)} + \rho_{11}^{c(ss)} = 1 $, we then obtain 
\begin{align}
\rho_{gg}^{c(ss)} = \frac{n_2}{2n_2+1},\quad \rho_{11}^{c(ss)} = \frac{n_2+1}{2n_2+1}.\label{rncs}
\end{align}
Similarly we obtain steady state solution for the thermal bath using Eq. (\ref{eq4}):
\begin{align}
\rho_{gg}^{th(ss)}=\frac{n_0+1}{2n_0+1},\quad \rho_{11}^{th(ss)}=\frac{n_0}{2n_0+1}.\label{rthss}
\end{align}
We then solve Eq.(\ref{eq1L}) non perturbatively over the $\Omega_p$ by factorizing Eqs. (1) into two uncoupled system of equations: one part containing the ground state population $\rho_{gg}$ and coherences $\rho_{g2}$ and $\rho_{2g}$ and another containing the excited state populations $\rho_{11}$ and $\rho_{22}$.  Since electronic coherence $\rho_{g2}$ evolves fast,  we can assume stationary population for the excited state $2$ while keeping the ground state evolution exactly. To have correct solution in the steady state we match the value of the stationary population for the excited state $2$ with its steady state solution: $\rho_{22}\to\rho_{22}^{'c(ss)}$ while initial conditions are $\rho_{gg}(0)=1$, $\rho_{g2}(0)=\rho_{2g}(0)=0$.
In this case solution for the coherences yields:
\begin{align}
\rho_{g2}^c(t)= \frac{4 i \left(2 n_2+1\right) \Omega_p  e^{-\frac{1}{4} \Gamma _2 \left(n_2+1\right) t} \sinh \left(\frac{1}{4} t \mathcal{Z}\right)}{\left(3 n_2+1\right) \mathcal{Z}},\label{rtg2}
\end{align}
where $\mathcal{Z} = \sqrt{\Gamma _2^2 \left(n_2+1\right){}^2-32 \Omega_p ^2}$. while the ground state population reads (assuming $4\sqrt{2}\Omega_p\ll\Gamma_2(n_2+1)$
\begin{align}
\rho_{gg}^c(t)= \frac{n_2}{3 n_2+1} + \frac{\left(2 n_2+1\right)e^{-\tilde{\Gamma}t} }{3 n_2+1},\label{rtgg}
\end{align}
We now calculate the population of states $1$ and $2$ using the solutions for $\rho_{g2}$ and $\rho_{2g}$ given by Eq. (\ref{rtg2}), and initial conditions are $\rho_{11}(0) = \rho_{22}(0) = 0$ which yields
\begin{align}
\rho_{11}(t)= \frac{\left(n_2+1\right) \left(1-e^{-\tilde{\Gamma}t}\right)}{3 n_2+1}, \rho_{22}(t) =\frac{n_2 \left(1-e^{-\tilde{\Gamma}t} \right) }{3 n_2+1}.\label{rt11}
\end{align}\\

\setcounter{equation}{0}
\renewcommand{\theequation}{C\arabic{equation}}
\begin{center}
\textbf{\large{Appendix C: Three level QHE}}
\end{center}
A detailed derivation of the matrix equations for a three level atom and induced coherence given in the Supplementary of \cite{sd11}. Here we consider a three level system $g, 0$ and $1$ and the Hamiltonian for the three level system is given by
\begin{equation}
H_{0} = \sum_{i=g,0,1}^{}\omega_{i}|i\rangle\langle i|,
\end{equation}
while the probe-system interaction is described by \cite{kpra18}
\begin{equation}
V(t) = \lambda ( e^{i\omega t}|1\rangle\langle 0| + e^{-i\omega t} |0\rangle\langle 1|),
\end{equation}
where $\lambda$ is a probe-matter coupling. Transition $0-g$ is in contact with the cold reservoir at temperature $T_c$. The corresponding Liouville operator for the cold bath-system interaction is given by Eq. (\ref{s9}).

Effective hot bath  at temperature $T_h$ is driving $g-1$ transition described by the Liouville operator
\begin{eqnarray}
\mathcal{L}_{h}\left[\rho\right] &=&\nonumber \Gamma_{h}(n_{h} + 1)\left[2|g\rangle \langle g|\rho_{11} - |1\rangle\langle 1|\rho - \rho |1\rangle\langle 1|\right]\\ &+& \Gamma_{h}n_{h}\left[2|1\rangle\langle 1|\rho _{gg} - |g\rangle\langle g|\rho - \rho |g\rangle\langle g| \right].
\end{eqnarray}
Here the average occupation numbers are given by
\begin{equation}
n_{c} = (e^{\frac{\omega_{c}}{T_{c}}} - 1)^{-1}, n_{h} = (e^{\frac{\omega_{h}} {T_{h}}} - 1)^{-1}.
\end{equation}
The evolution of the density matrix reads
\begin{eqnarray}
\dot{\varrho_{00}} &=& -\nonumber i\lambda(\rho_{10}-\rho_{01}) - 2\Gamma_{c}(1 + n_{c})\rho_{00} + 2\Gamma_{c}n_{c}\rho_{gg},\\
\dot{\varrho_{11}} &=&\nonumber  i\lambda(\rho_{10}-\rho_{01}) - 2\Gamma_{h}(1 + n_{h})\rho_{11} + 2\Gamma_{h} n_{h} \rho_{gg},\\\nonumber
\dot{\varrho_{01}} &=&\nonumber -i\Delta\rho_{01} -  i\lambda(\rho_{11}-\rho_{00}) - \Gamma_{c}(1 + n_{c})\rho_{01}\\
&-&\nonumber \Gamma_{h}(1 + n_{h})\rho_{01}, \nonumber\\
\dot{\varrho_{10}} &=&\nonumber i\Delta\rho_{01} +  i\lambda(\rho_{11}-\rho_{00}) - \Gamma_{c}(1 + n_{c})\rho_{10}\\
&-& \Gamma_{h}(1 + n_{h})\rho_{10}, \label{eomb}
\end{eqnarray}
where $\Delta = \omega + \omega_{0}- \omega_{1} $ is the detuning between the output laser radiation and molecular level transition; $ n_{h} $ and $n_{c}$, $\Gamma_{h} $ and $\Gamma_{c}$ are the average occupation numbers and dephasing rates for the hot and cold baths, respectively.\\
Solving Eqs.(\ref{eomb}) in the steady state assuming $\Delta=0$ we obtain for the power  and efficiency defined by Eq. (\ref{eq:PQeta}) as well as the heat flux\cite{kpra18}
\begin{align}\label{eq:Pdef}
&P= i\hbar\lambda(\omega_{c}-\omega_{h})(\varrho_{01}-\varrho_{10}),\notag\\
&\dot{Q}_{h}= i\hbar\omega_{h}\lambda(\varrho_{01}-\varrho_{10}), \notag\\
&\eta = 1 - \frac{\omega_{c}}{\omega_{h}}.
\end{align}
where $\omega_{h} = \omega_{1} - \omega_{g}$ and $\omega_{c} = \omega_{0} - \omega_{g}$. After solving Eq. (\ref{eomb}) the output power reads
\begin{equation}
P = \frac{2}{3} \frac{\lambda^{2}\Gamma_{h}\Gamma_{c}(n_{c} - n_{h}) (\omega_{c} - \omega_{h}) }{(\Gamma_{h}n_{h} + \Gamma_{c}n_{c}) (\lambda^{2} + \Gamma_{h}\Gamma_{c}n_{c}n_{h} ) }.\label{genp}
\end{equation}
In the high temperature limit $n_{c} = n_{1} \simeq T_{c}/ \omega_{c}$, $n_{2} \simeq T_{c}/\omega_{21}$ (both $n_{2}$ and $n_{c}$ have the same "cold" temperature since they are related to the same ambient (phonon) environment). Introducing dimensionless parameters: $\omega_{h} = c\omega_{c}$ and $c_{21} = (c_p - c)\omega_{c}$ where $c_p = \omega_{p}/\omega_{c}$ and $c_{21} = \omega_{21}/ \omega_{c}$, the effective hot bath temperature $T_{h} = \omega_{c}\sqrt{2\Omega^{2}/\gamma_{2} \Gamma_{c}}$ Eq. (\ref{genp}) reads
\begin{equation}
\mathcal{P} = \frac{2P_0  \lambda^{'2} c_{21} (c_p-1)(1-c_{21})}{3 \left(\lambda^{'2}+c_{21}\right) \left(\tau^2 + c_{21}\right)},\label{gpdl}
\end{equation}
where $P_0=\Gamma_c\omega_c$, $\lambda'=\lambda\omega_c/(\Gamma_cT_h)$, $\tau = T_{c}/T_{h}$ and we assume that $c_p\gg c_{21}$ valid for e.g. visible range for the pump field and IR phonon range. Using the above dimensionless parameters we recast the efficiency in Eq. (\ref{eq:Pdef}) as
\begin{align}
\eta=1-\frac{1}{c_p-c_{21}}.\label{effa}
\end{align}
Further maximization of Eq. (\ref{gpdl}) with respect to $c_{21}$ yields Eq. (\ref{gnpm}). The corresponding efficiency at maximum power is given by Eq. (\ref{effg}). \\

\setcounter{equation}{0}
\setcounter{figure}{0}
\renewcommand{\theequation}{D\arabic{equation}}
\renewcommand{\thefigure}{D\arabic{figure}}
\begin{center}
\textbf{\large{Appendix D: Perturbative pump-probe signal}}\end{center}
\begin{figure}[t]
	\includegraphics[width=\linewidth]{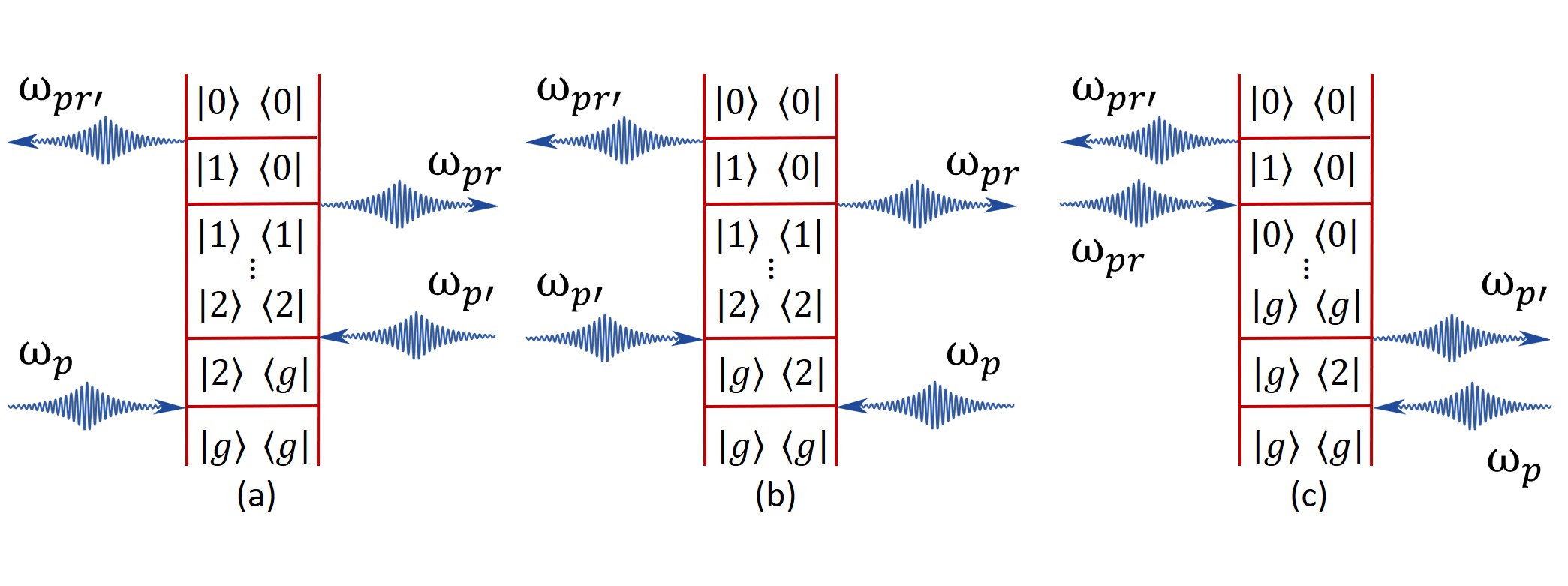}
	\caption{(color online) The double sided Feynman-diagram representing the pump-probe signal. There are total six pathways that contribute to the pump-probe signal (diagrams a, b, c and their complex conjugate).  }\label{fey}
\end{figure}
The pump-probe signal is given by Eq. (\ref{eq:Ps}).
The response function can be read off the diagrams in Fig. \ref{fey}, which yields
\begin{widetext}
\begin{align}
\mathcal{R}_{a}(-\omega_{pr'},&\omega_{pr},-\omega_{p'},\omega_{p}) =\notag\\
&=\langle \mathcal{V}_{00,10} \mathcal{G}_{10,10}({\omega_{pr}}+\omega_{p} - \omega_{p'})\mathcal{V}_{10,11} \mathcal{G}_{11,22}(\omega_{p} - \omega_{p'})
\mathcal{V}_{22,2g} \mathcal{G}_{2g,2g} ({\omega_{p}}) \mathcal{V}_{2g,gg}\rangle,
\label{R1}
\end{align}
\begin{align}
\mathcal{R}_{b}(-\omega_{pr'},&\omega_{pr},\omega_{p'},-\omega_{p}) =\notag\\
&=\langle \mathcal{V}_{00,10} \mathcal{G}_{10,10}({\omega_{pr}}-\omega_{p} + \omega_{p'})\mathcal{V}_{10,11} \mathcal{G}_{11,22}(-\omega_{p} + \omega_{p'})
\mathcal{V}_{22,g2} \mathcal{G}_{g2,g2} (-{\omega_{p}}) \mathcal{V}_{g2,gg}\rangle,
\label{R2}
\end{align}
\begin{align}
\mathcal{R}_{c}(-\omega_{pr'},&\omega_{pr},\omega_{p'},-\omega_{p}) =\notag\\
&=\langle \mathcal{V}_{00,10} \mathcal{G}_{10,10}({\omega_{pr}}-\omega_{p} + \omega_{p'})\mathcal{V}_{10,00} \mathcal{G}_{00,gg}(-\omega_{p} + \omega_{p'})
\mathcal{V}_{gg,g2} \mathcal{G}_{g2,g2} (-{\omega_{p}}) \mathcal{V}_{g2,gg}\rangle.
\label{R3}
\end{align}
\end{widetext}
Perturbative result in Eqs.(\ref{R1}) - (\ref{R3}) contains both population and coherence Green's functions. The coherence Green's functions for 1-0 transition is given by $\mathcal{G}_{10,10} (\omega)= -[i(\omega - \omega_{10}) - \Gamma_{10}]^{-1} $, for $g$-2 transition $ \mathcal{G}_{2g,2g} (\omega) = -[  i \left(\omega-\omega _{2 g}\right) - \Gamma_{2g}]^{-1}$, where $\Gamma_{10} =[\Gamma_{c}(n_{c} + 1) + \Gamma_{2}n_{2}]/2$ and $\Gamma_{2g}= [\Gamma_{c}n_{c} + \Gamma_{2}(n_{2} + 1)]/2$. The population Green's function is a solution of the coupled transport (relaxation) equations:
\begin{eqnarray}
\dot{\rho_{22}} &=&\nonumber -\Gamma_{2} (n_{2}+1)\rho_{22} + \Gamma_{2} n_{2} \rho_{11},\\
\dot{\rho_{11}} &=& \Gamma_{2} (n_{2}+1)\rho_{22} - \Gamma_{2} n_{2} \rho_{11},
\label{eq.1}
\end{eqnarray}
\begin{eqnarray}
\dot{\rho_{00}} &=& \nonumber-\Gamma_{c} (n_{c}+1)\rho_{00} + \Gamma_{c} n_{c} \rho_{gg},\\
\dot{\rho_{gg}} &=& \Gamma_{c} (n_{c}+1)\rho_{00} - \Gamma_{c} n_{c} \rho_{gg}.
\label{eq.1a}
\end{eqnarray}
Eqs. (\ref{eq.1}) - (\ref{eq.1a}) can be recast as a Pauli master equation:
\begin{eqnarray}
\dot{\rho}_{ii}(t) = -\sum_{ii,jj}^{}\kappa_{ii,jj} \rho_{jj}(t),
\label{eq.2p}
\end{eqnarray}
where, $\kappa_{ii,jj}$ is the population transport matrix.  In Eq. \ref{eq.2p}, the diagonal elements, $i=j$, $\kappa_{ii,ii}$ are positive, whereas the off-diagonal elements, $i \neq j$, $\kappa_{ii,jj}$ are negative. The population transport matrix satisfies the population conservation:   $\sum_{i}^{}\kappa_{ii,jj}=0$. The evolution of the diagonal elements is defined by the population Green function, $\rho_{jj}(t) = \sum_{i}\mathcal{G}_{jj,ii}(t)\rho_{ii}(0)$. where $\mathcal{G}_{jj,ii}(t)$ is given \cite{mcr9}
\begin{eqnarray}
\mathcal{G}_{jj,ii}(t) = \sum_{n}^{}\xi_{jn}^{(R)} D_{nn}^{-1}\exp(-\lambda_{n}t) \xi_{ni}^{(L)},
\label{eq.2g}
\end{eqnarray}
where $\lambda_{n}$ is the $nth$ eigenvalue of left and right eigenvector $(\xi_{n}^{(L)},\xi_{n}^{(R)})$ and $D = \xi^{L} \xi_{R}$ is a diagonal matrix. Using Eq. (\ref{eq.2g}) we obtain for the population Green's functions:
\begin{eqnarray}
\mathcal{G}_{00,gg}(t) &=& \frac{n_c(1- e^{-t(1+2 n_{c})\Gamma_{c} })  }{(1+2  n_{c})}\label{g0},\\
\mathcal{G}_{11,22}(t) &=& \frac{(1+n_2)(1- e^{-t(1+2 n_{2})\Gamma_{2} })  }{(1+2  n_{2})}.
\label{g12}
\end{eqnarray}
Assuming narrowband resonant pump and probe fields $\omega_{p'} = \omega_{p}=\omega_{2g}$ and $\omega_{pr'} = \omega_{pr}=\omega_{10}$ Eqs. (\ref{R1}) - (\ref{R3}) read
\begin{align}
\mathcal{R}_{a}=\mathcal{R}_b=\frac{4|\mu_{10}|^2|\mu_{2g}|^2(1+n_2)}{(n_c\Gamma_c+n_2\Gamma_2)^2(1+2n_2)\sigma_p},
\end{align}
\begin{align}
\mathcal{R}_{c}=\frac{4n_c|\mu_{10}|^2|\mu_{2g}|^2}{(n_c\Gamma_c+n_2\Gamma_2)^2(1+2n_c)\sigma_p},
\end{align}
where $\sigma_p$ is an infinitesimal parameter of the order of the pump bandwidth required for the convergence of the Fourier transform. Taking high temperature limit $n_c\gg 1$, $n_2\gg 1$ the total power (\ref{eq:Ps}) reads
\begin{align}
P^s=\frac{12(\omega_h-\omega_c)\lambda^2\Omega_p^2}{(n_c\Gamma_c+n_2\Gamma_2)^2\sigma_p}, \label{d12}
\end{align}
where we used $\omega_{10}=\omega_h-\omega_c$, $\lambda=\mu_{10}\mathcal{E}_{pr}$, $\Omega_p=\mu_{2g}\mathcal{E}_p$. Expanding occupation numbers in the high temperature limit $n_c=T_c/\omega_c$, $n_2=T_c/\omega_{21}$ and maximizing the power with respect to $c_{21}=\omega_{21}/\omega_c$ we obtain Eq. (\ref{eq:Pmaxs}).


\end{document}